# Dynamics of Exozodiacal Clouds

Marc Kuchner, Christopher Stark, NASA GSFC
Olivier Absil and Jean-Charles Augereau, Grenoble Observatory
and Philippe Thébault, Paris Observatory

The inner Solar System contains a cloud of small (1-100 µm) dust grains created when small bodies---asteroids, comets, and Kuiper belt objects---collide and outgas. This dust cloud, the *zodiacal* cloud probably has extrasolar analogs, *exozodiacal* clouds. Exozodiacal clouds are related to *debris disks,* clouds of rocks and dust orbiting main sequence stars thought to represent the debris left over from planet formation. Some debris disks appear to contain distinct inner clouds that could be considered massive exozodiacal clouds (e.g. Koerner et al. 1998, Absil et al. 2006).

This white paper addresses the need for future theoretical work on the dynamics of exozodiacal clouds. This theoretical work should help us discover new planets and understand exozodiacal clouds as astrophysical noise. So far, observations of nearby stars have not provided good constraints on the structures of exozodiacal clouds. But future observations probably will demand a better theoretical understanding of these systems.

**Recognizing Planets Via Disk Structures:**

As zodiacal dust particles spiral inwards under the effect of Poynting-Robertson drag, they interact dynamically with the planets, and sometimes become temporarily trapped in mean motion resonances (MMRs) with planets. This resonant trapping can produce a detectable density structure in the dust cloud. The solar zodiacal cloud shows evidence of this kind of pattern and other structures created by planet – dust interactions. It also bears the signatures of individual comets and individual collisional cascades in the asteroid belt.

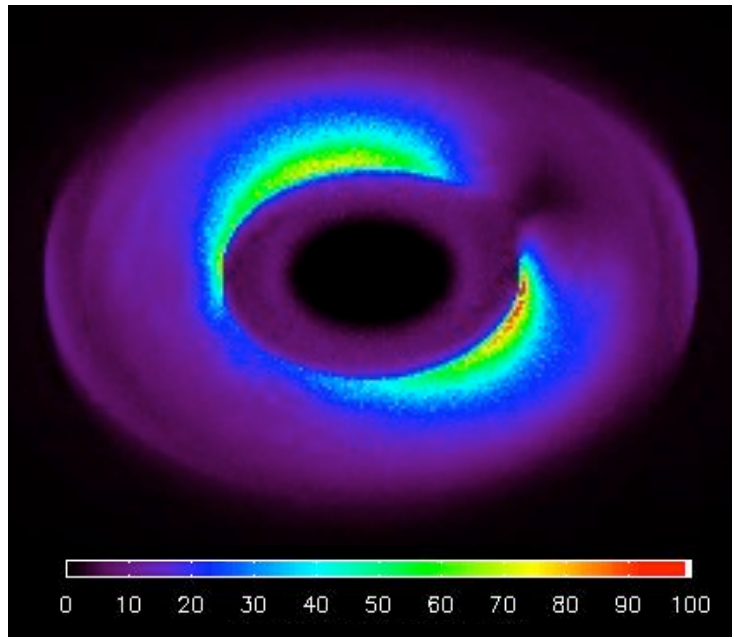

**Figure 1: Ring of dust trapped in mean motion resonances with a 5 Earth-mass planet (Stark & Kuchner 2007, in preparation).**

Figure 1 shows a surface density map for a simulated dust cloud containing a 5 Earth-mass planet computed using the Thunderhead supercomputer at Goddard (Stark & Kuchner 2007, in preparation). The map shows a torus of dust trapped in MMRs with the planet. Recognizing these structures in images of exozodiacal clouds and debris disks can allow us to indirectly detect extrasolar planets that otherwise would go unseen and measure their masses and orbital parameters. This method may be the only way to detect extrasolar planets in long orbital periods (e.g. ~100 years) that are not self-luminous.

Recognizing planets this way in massive debris disks (like the disk around β Pictoris) can be fraught with degeneracies and modeling uncertainties. However, upcoming telescopes like LBTI (the Large Binocular Telescope Interferometer), ALMA, JWST and many proposed missions like TPF, Eclipse, EPIC, FKSI, SPIRIT, etc. have the potential to find debris disks with much less optical depth than the ones known now.

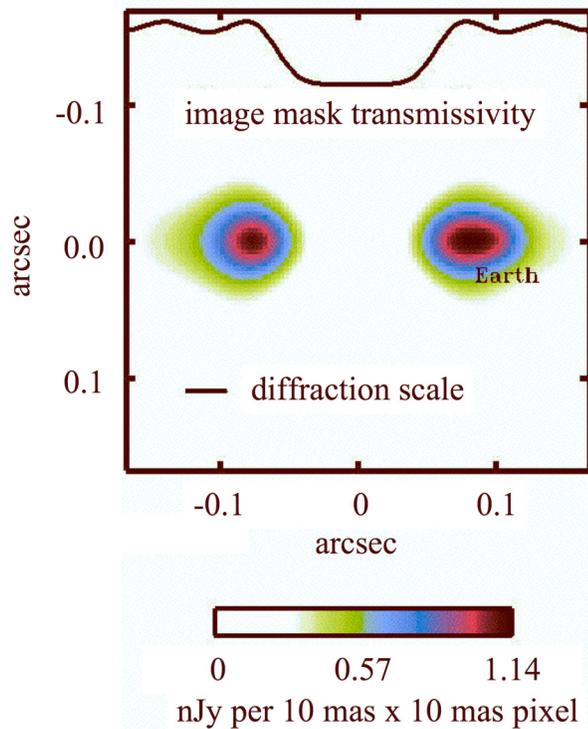

Figure 2: Simulated image of a Solar System analog (minus Venus) as seen by baseline TPF-C at 10 pc. Most of the signal is exozodiacal light. (Kuchner)

These sparse disks should have physics that is relatively simple and similar to that of the solar zodiacal cloud, allowing us to decode their structures straightforwardly, if our theoretical progress has prepared us.

**Dust Patterns as Astrophysical Noise:**

Exozodical dust is important as a potential source of astrophysical noise for planet finding; this concern also drives modeling efforts.

The capability of imaging extrasolar planets carries with it the capability of imaging exozodiacal clouds. For example, the solar zodiacal cloud is the brightest component of the solar system in the optical and infrared except for the Sun. In a sense, TPF might be appropriately called the Exozodiacal Cloud Finder.

Figure 1 shows a view of how the solar system (minus Venus) would appear to the current baseline TPF-C (Terrestrial Planet Finder-Coronagraph) design. Most of the light in the image comes from exozodiacal dust. The Earth appears as a difference between the shapes of the two lobes of the image.

As Figure 1 suggests, ~10% surface brightness fluctuations in a bright exozodiacal cloud on the scale of ~0.1 AU can masquerade as terrestrial planets to TPF; they become confusion noise. Photon noise in the light from exozodiacal clouds limits the spectroscopic capability of planet-imaging telescopes like TPF. We need better tmodels to predict the power level in these noise terms.

## Work Needed

**Quantitative Models of Surface Brightness and Contrast:**

Models of steady state zodiacal cloud dynamics have come a long way since Dermott et al (1994) first recognized the leading/trailinig asymmetry in the surface brightness of the infrared sky as a circumsolar ring of zodiacal dust trapped in MMRs with the Earth. We now more-or-less understand the geometry of resonant signatures from analytical calculations (Kuchner & Holman 2002) and numerical simulations (e.g. Deller & Maddison 2005). Figure 3 shows some of the basic density wave patterns associated with various MMRs (Kuchner & Holman 2002).

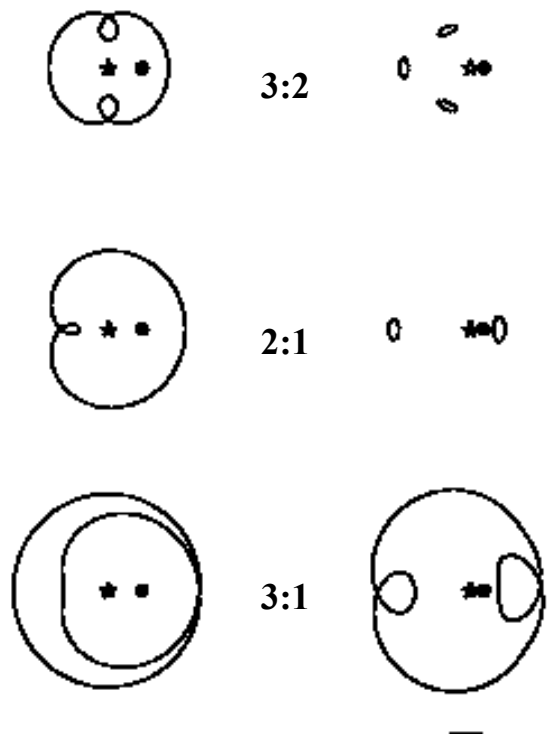

However, although the geometry of the resonances has become reasonably well understood, the way resonances are populated is hard to model. In a way, modeling exozodiacal clouds is like modeling the spectra of stars; it's easy to understand where the lines (the resonances) are, but it's harder to model how strong the lines (how populated the resonances) are. This uncertainty limits our ability to model the surface brightness perturbations created by planets in dust clouds.

Figure 3 shows the results of a simulation by Moro-Martin and Malhotra (2003) showing the semi-major axis distribution of dust interacting with a Neptune-like planet. The three curves show the results of three simulations of 100 particles each. The differences between the curves show that the three simulations populated the MMRs differently, resulting in three different model images. This Figure suggests that future simulations will need thousands of particles in each size bin to reduce this Poisson noise to the level where they can produce reliable

**Figure 3:** Density wave patterns associated with some of the key MMRs. The stars are the stars; the filled circles are the planets. Patterns associated with low-eccentricity planets are on the left; patterns for high-eccentricity planets are on the right. (Kuchner & Holman 2002)

quantitative predictions of the contrast in any resonant structure. This computational power is within easy reach of today's large computer clusters.

**Multiple-Planet Systems**

Many extrasolar planet systems contain multiple planets. So far, the only multiple-planet system that has been examined in detail—including planet-planet interactions---is the Solar System. Multiple planets present a special numerical challenge because they add another time scale to the problem: the time scale of secular orbital evolution. Handling the secular dynamical effects will requires more algorithm development and more computational power.

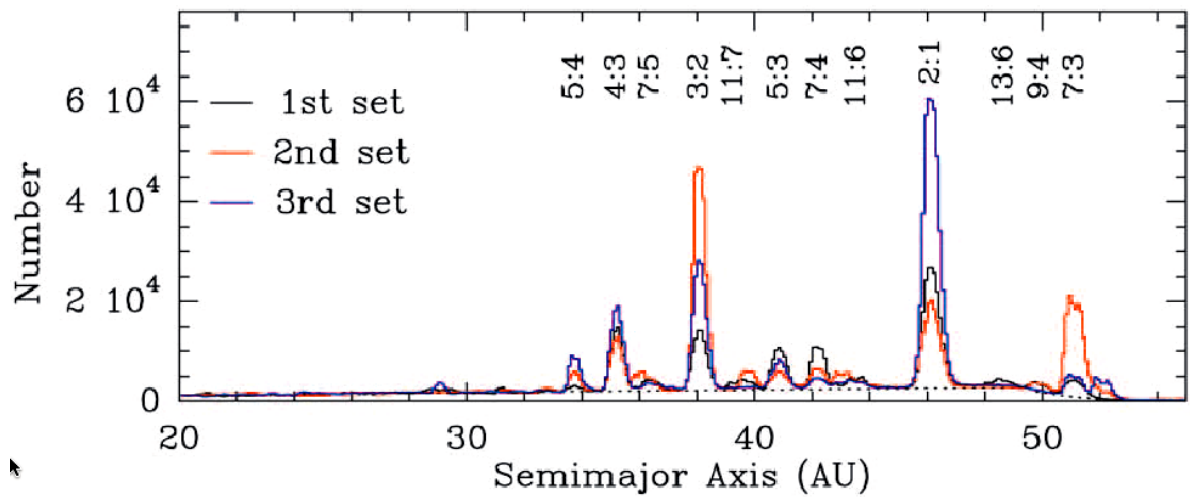

Figure 4: Semi-major axis distribution for 3 simulations of resonant trapping. The different simulations show different populations of the major resonances (Moro - Martin & Malhotra 2003).

**Collisional Effects and Gas Drag**

For exozodiacal clouds with optical depths >50 times that of the solar zodiacal cloud, grain-grain collisions become important. There have been some efforts to model what this phenomenon does to the three-dimensional distribution of dust in these clouds (e.g. Kehoe & Dermott 2005). However, the problem of modeling collisions in exozodiacal dust clouds is fundamentally challenging because the number of particles is large ($10^{30}$) and the collisional cross-sections are small ($10^{-32}$ $AU^2$), and it has never been treated in a self-consistent manner. Understanding collisional effects is crucial to understanding the dynamics of most currently known debris disks, which have optical depths 100-10,000 times that of the solar zodiacal cloud. This problem may yield to a statistical treatment (see Thébault et al. 2003, Krivov et al. 2006).

In these massive disks, gas drag can also become important. Recent spectroscopic measurements show that some debris disks contain dynamically-significant quantities of atomic and molecular gas, perhaps released during collisions (e.g. Roberge et al. 2006, Chen et al. 2006). Thebault & Augereau (2005) and others have modeled the consequences of gas drag on the one-dimensional structure of circumstellar clouds, but the three-dimensional effects have not yet been studied.

**Observational Tests**

Presently, there is only one known example of a dust cloud structure created by a planet where we know from independent observations what the mass and orbit of the planet are: the Earth's circumsolar ring. This ring has been observed by IRAS and COBE and is now being observed by Spitzer. The geometry of this ring is somewhat well understood, but the contrast between this ring and the background clouds is not! We need to find and model more examples of this phenomenon and we need better observations of the solar system and other planetary systems to build confidence in the models.